\documentclass[12pt]{article}
\usepackage{amscd,amsmath,amssymb,amstext,amsthm,exscale,latexsym}
\usepackage{graphicx,floatflt}
\newcommand {\al}   {\alpha}       \newcommand {\bt}  {\beta}
\newcommand {\g }   {\gamma}       \newcommand {\G }  {\Gamma}
\newcommand {\dl}   {\delta}

\newcommand {\lm}   {\lambda}      
          
\newcommand {\s }   {\sigma}       
         
\newcommand {\vf }  {\varphi}      
         \newcommand {\om}  {\omega}

\newcommand {\pl}   {\partial}     \newcommand {\nb}  {\nabla}

\newcommand   {\const}{{\sf\,const}}
\newcommand   {\diag}{{\sf\,diag\,}}

\newcommand {\vol}  {\sqrt{|g|}}
\newcommand {\Sa}  {{\textsc{a}}}

\newcommand {\MM}  {{\mathbb M}}
\newcommand {\MS}  {{\mathbb S}}
\newcommand {\MO}  {{\mathbb O}}

\newcommand {\MR}  {{\mathbb R}}
\newcommand {\CC }  {{\cal C}}
      
\newcommand {\Go}  {\mathfrak{o}}   \newcommand {\Gs}  {\mathfrak{s}}
\begin{document}
\title     {Tube dislocations in gravity}
\author    {G. de Berredo-Peixoto
           \thanks{E-mail: guilherme@fisica.ufjf.br}\\ \\
           \sl Departamento de Fisica, Universidade Federal de Juiz de Fora,\\
           \sl Juiz de Fora. CEP 36036--330, MG, Brazil\\ \\
           M. O. Katanaev
           \thanks{E-mail: katanaev@mi.ras.ru}\\ \\
           \sl Steklov Mathematical Institute,\\
           \sl Gubkin St. 8, Moscow, 119991, Russia}
\date      {29 September 2008}
\maketitle
\begin{abstract}
We consider static massive thin cylindrical shells (tubes) as the sources in
Einstein's equations. They correspond to $\dl$- and $\dl'$-function type
energy-momentum tensors. The corresponding metric components are found explicitly.
They are not continuous functions in general and lead to ambiguous curvature
tensor components. Nevertheless all ambiguous terms in Einstein's equations
safely cancel. The interplay between elasticity theory, geometric theory of
defects, and General Relativity is analyzed. The elasticity theory provides a
simple picture for defects creation and a new look on General Relativity.
\end{abstract}
\vskip10mm
\section{Introduction}
In gravity models, metric components are usually considered as continuous and
two times differentiable functions. This property is needed for unambiguous
calculation of curvature tensor components. In a general case, if metric
components are not continuous\footnote{We do not consider the case when
discontinuity of metric components arises from discontinuous coordinate
transformations for smooth curvature on a manifold. Instead, we are stick to the
case when discontinuity of metric components is produced by discontinuity of
curvature.},
then Christoffel's symbols contain $\dl$-function, and curvature tensor will
include squares of $\dl$-functions, which cannot be unambiguously defined.
However, there may exist rare particular cases when all ambiguous terms in
Einstein's equations cancel. This means that if we regularize the corresponding
discontinuous metric components by smooth functions, then the solution of
Einstein's equations will not depend on a regularization. This situation is
acceptable and worth to be analyzed. It may happen when Einstein's equations
reduce to one or a system of {\em linear} differential equations for a specific
combinations of metric components. For example, it is well known that Einstein's
equations reduce to the linear inhomogeneous Poisson equation for the conformal
factor of the two dimensional part of the metric for a cosmic string
\cite{DeJatH84}. The corresponding metric components (or its inverse) in
Cartesian coordinates on a plane are singular at the location of conical
singularity.

In the present paper, we consider tube dislocations in space-time which
correspond to a matter distributed uniformly along static straight circular
tubes which are infinitely thin. There are two types of tube defects with the
energy-momentum tensor proportional to the $\dl$-function or its derivative,
respectively. In the first case, the space-time has conical singularity either
inside or outside the tube, while in the second case, the space-time has no
conical singularities.

Thin shells are well known in General Relativity for continuous metric
components \cite{MiThWh73}. For these types of metrics, Christoffel's symbols
and the curvature tensor may have jumps and $\dl$-functions, respectively. Then
Einstein's equations yield the matching conditions on the shell in terms of
extrinsic curvature. We consider a different situation when metric components
have jumps but all ambiguous terms in Einstein's equations cancel.

We start the analysis and derive the metric for the tube dislocation within the
static linear elasticity theory because this approach clarifies a lot the whole
construction. We show that the metric for a tube dislocation satisfies
Einstein's equations with the $\dl'$ source. Then we consider this problem
within the geometric theory of defects developed in
\cite{KatVol92,KatVol99,Katana03,Katana04} (for review see \cite{Katana05}, the
combined wedge and edge dislocation was considered in \cite{deBKat07}) and
generalize the metric to four dimensions. Afterwards, we consider conical tube
dislocation and asymptotically flat wedge dislocation. The first one is flat in
the inside region and conical outside, whereas the second is conical inside and
flat outside. These dislocations have a $\dl$-function source. At the end, we
consider an arbitrary continuous distribution of tube dislocations.
\section{Tube dislocation in the linear elasticity theory}
We consider a three-dimensional Euclidean space $\MR^3$ (infinite homogeneous
and isotropic elastic media or eather in general relativity) with Cartesian
coordinates $x^i,y^i$, $i=1,2,3$. The Euclidean metric is denoted by
$\dl_{ij}=\diag(+++)$. The basic variable in the elasticity theory is
the displacement vector field $u^i(x)$, $x\in\MR^3$, which measures the
displacement of a point in the elastic media. In the absence of external forces,
Newton's and Hook's laws reduce to three second order partial differential
equations which describe the equilibrium state of elastic media
(see, i.e.\ \cite{LanLif70}),
\begin{equation}                                                  \label{eqsepu}
  (1-2\s)\triangle u_i+\pl_i\pl_j u^j=0,
\end{equation}
where
\begin{equation*}
  \s=\frac\lm{2(\lm+\mu)}
\end{equation*}
is the dimensionless Poisson ratio, ($-1\le\s\le1/2$), and $\triangle$ is the
Laplace operator. The Lame coefficients $\lm,\mu$ characterize the elastic
properties of media. Raising and lowering of Latin indices $i,j,\dotsc$ is
performed using the Euclidean metric $\dl_{ij}$ and its inverse $\dl^{ij}$.
The boundary conditions for Eq.(\ref{eqsepu}) correspond to the physical problem
which is to be solved.

\begin{figure}[h,t]
\hfill\includegraphics[width=\textwidth]{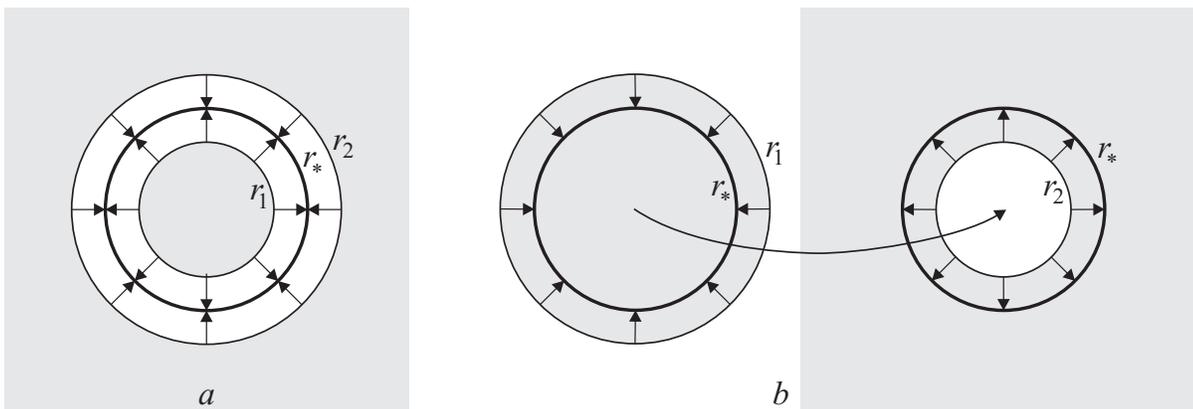}
\hfill {}
\\
\centering \caption{\label{ftubes} Negative ({\it a}) and positive ({\it b})
tube dislocations.}
\end{figure}

Let us pose the problem for the tube dislocation shown in Fig.\ref{ftubes},
{\it a}.
This dislocation is produced as follows. We cut out the thick cylinder of media
located between two parallel cylinders of radii $r_1$ and $r_2$ ($r_1<r_2$) with
axis $z=x^3$ as the axis of both cylinders, move symmetrically both cutting
surfaces one to the other and afterwards glue them. In the equilibrium state,
the gluing surface is also the cylinder, due to circular and translational
symmetries of the problem, of radius $r_*$ which is to be found.

In our conventions, Fig.\ref{ftubes},{\it a} shows the negative tube dislocation
because part of the media was removed, $r_1<r_2$. This procedure can be inverted
by addition of extra media to $\MR^3$ as shown in Fig.\ref{ftubes},{\it b}. In
this case, we call it positive tube dislocation, $r_1>r_2$.

This problem is naturally formulated and solved in cylindrical coordinates
$r,\vf,z$. Let us denote the displacement field components in cylindrical
coordinates by $u^r,u^\vf,u^z$. In our case, $u^\vf=0$, $u^z=0$, due to the
symmetry of the problem, and the radial displacement field, $u^r(r)$, depends
only on the radius $r$, and we drop the index, $u^r(r)=u(r)$, for simplicity.

The boundary conditions for the tube dislocation are
\begin{equation}                                                  \label{ebotud}
  u|_{r=0}=0,~~~~u|_{r=\infty}=0,~~~~
  \left.\frac{du_\text{in}}{dr}\right|_{r=r_*}
  =\left.\frac{du_\text{ex}}{dr}\right|_{r=r_*}.
\end{equation}
The first two conditions are purely geometrical, and the third one means the
equality of normal elastic forces inside and outside the gluing surface in the
equilibrium state. The subscripts ``in'' and ``ex'' denote the displacement
vector field inside and outside the gluing surface, respectively.

Our definition of the displacement vector field differs slightly from the usual
one. In our notations, the point with coordinates $y^i$ after elastic
deformation moves to the point with coordinates $x^i$:
\begin{equation}                                                  \label{eeldef}
  y^i\rightarrow x^i(y)=y^i+u^i(x),
\end{equation}
the displacement vector field being the difference between new and old
coordinates, $u^i(x)=x^i-y^i$ (this is usual). The difference is that we
consider the components of the displacement vector field $u^i(x)$ as functions
of the final state coordinates of media points $x^i$ and not of the initial ones
$y^i$. This is equivalent to the standard approach in the absence of
dislocations because both sets of coordinates $x^i$ and $y^i$ cover the entire
Euclidean space $\MR^3$. If the dislocation is present, the final state
coordinates $x^i$ cover the whole $\MR^3$ while the initial state coordinates
cover only part of the Euclidean space lying outside the thick cylinder which
was removed. Therefore the final state coordinates are preferable in the
presence of dislocations. This difference can be considered as inessential in
the linear approximation but the geometric theory of defects describes large
deformations along with the small ones.

The elasticity equations (\ref{eqsepu}) can be easily solved in the considered
case of tube dislocation. The Laplacian of the radial component, which is the
only one that differs from zero, and the divergence in the cylindrical
coordinates have the form:
\begin{align*}
  \triangle u_r&=\frac1r\pl_r(r\pl_r u_r)+\frac1{r^2}\pl^2_\vf u_r+\pl^2_z u_r
  -\frac1{r^2}u_r-\frac2{r^2}\pl_\vf u_\vf,
\\
  \pl_i u^i&=\frac1r\pl_r(ru^r)+\frac1r\pl_\vf u^\vf+\pl_z u^z,
\end{align*}
where the indices are lowered using the Euclidean metric in cylindrical
coordinates: $u_r=u^r$, $u_\vf=u^\vf r^2$, and $u_z=u^z$. The angular $\vf$ and
$z$ components of equations (\ref{eqsepu}) are identically satisfied, and the
radial component reduces to the ordinary differential equation,
\begin{equation}                                                  \label{eradec}
  \pl_r\left[\frac1r\pl_r(r u)\right]=0,
\end{equation}
which has a general solution
\begin{equation*}
  u=ar-\frac br,~~~~a,b=\const,
\end{equation*}
depending on two arbitrary constants of integration $a$ and $b$. Due to the
first two boundary conditions (\ref{ebotud}), the solutions inside and outside
the gluing surface are
\begin{equation}                                                  \label{elsopt}
\begin{aligned}
  u_\text{in}&=ar, &a&>0,
\\
  u_\text{ex}&=-\frac br, &b&>0.
\end{aligned}
\end{equation}
The signs of the integration constants correspond to the negative tube
dislocation shown in Fig.\ref{ftubes},{\it a}. For positive tube dislocation,
Fig.\ref{ftubes},{\it b}, the integration constants have different signs, $a<0$
and $b<0$. The third boundary condition (\ref{ebotud}) determines the radius of
the gluing surface,
\begin{equation}                                                  \label{eglsur}
  r_*^2=\frac ba.
\end{equation}
After simple algebra the integration constants can be expressed through the
radii
\begin{equation}                                                  \label{eintco}
  a=\frac{r_2-r_1}{r_2+r_1}=\frac l{2r_*},~~~~b=\frac{r_2^2-r_1^2}4=\frac{lr_*}2,
\end{equation}
where
\begin{equation*}
  l=r_2-r_1,~~~~r_*=\frac{r_2+r_1}2
\end{equation*}
are the thickness of the removed cylinder and the radius of the gluing surface,
respectively. The first expression in (\ref{eintco}) restricts the range of
the integration constant, $0<|a|<1$. For negative and positive tube dislocations,
$l>0$ and $l<0$, respectively. We see that the gluing surface lies exactly
in the middle between radii $r_1$ and $r_2$. So, Eq.(\ref{elsopt}) with
integration constants (\ref{eintco}) yields a complete solution for the tube
dislocation within the linear elasticity theory. We expect it to be valid for
small relative displacements: $l/r_1\ll 1$, $l/r_2\ll1$.

Note that the answer within the linear elasticity theory does not depend on
the Poisson ratio of the media. In this sense, the tube dislocation is a purely
geometric defect which does not feel the elastic properties.

Now we compute the geometric quantities of the manifold corresponding to the
tube dislocation. From the geometric point of view, the elastic deformation
(\ref{eeldef}) is a diffeomorphism between given domains in the Euclidean space.
The original elastic media $\MR^3$, before the dislocation is made, is described
by Cartesian coordinates $y^i$ with the Euclidean metric $\dl_{ij}$. The pull
back of a diffeomorphism $x\rightarrow y$ induces nontrivial metric on $\MR^3$
corresponding to the tube dislocation. In Cartesian coordinates it has the form
\begin{equation*}
  g_{ij}(x)=\frac{\pl y^k}{\pl x^i}\frac{\pl y^l}{\pl x^j}\dl_{kl}.
\end{equation*}
We use curvilinear cylindrical coordinates for the tube dislocation and
therefore modify our notations. Let us denote indices in curvilinear coordinates
in the Euclidean space $\MR^3$ by Greek letters $x^\mu$, $\mu=1,2,3$. Then the
``induced'' metric for the tube dislocation in cylindrical coordinates
is\footnote{We put the word ``induced'' in inverted commas because on the
cutting surface the induced metric is not defined.}
\begin{equation}                                                  \label{eincym}
  g_{\mu\nu}(x)=\frac{\pl y^\rho}{\pl x^\mu}\frac{\pl y^\s}{\pl x^\nu}
  \overset{\circ}g_{\rho\s}
\end{equation}
where $\overset{\circ}g_{\rho\s}$ is the Euclidean metric written in cylindrical
coordinates. We denote cylindrical coordinates of a point before the dislocation
is made by $\lbrace y,\vf,z\rbrace$, where $y$ without any index stands for the
radial coordinate and we take into account that angular $\vf$ and $z$
coordinates are not changed. Then the diffeomorphism is described by a single
function relating old and new radial coordinates of a point $y=r-u(r)$, where
\begin{equation}                                                  \label{edifun}
  u(r)=\begin{cases} ar, &r<r_*, \\ -\displaystyle\frac br, & r>r_*. \end{cases}
\end{equation}
This function has the jump $ar_*+b/r_*=l$ across the cut and hence is not
continuous. Therefore a special care must be taken in calculation of the induced
metric components. To this end we introduce the function
\begin{equation}                                                  \label{edefiu}
  v=\begin{cases} a, & r\le r_*, \\ \displaystyle\frac b{r^2}, & r\ge r_*,
\end{cases}
\end{equation}
which is continuous across the cutting surface. This function differs from the
derivative with respect to $r$ of the displacement vector field $u(r)$ defined
in (\ref{edifun}) by the $\dl$-function
\begin{equation}                                                  \label{ereluv}
  u'=v-l\dl(r-r_*).
\end{equation}
The induced metric outside the cut is given by expression (\ref{eincym}) and is
not defined on the cutting surface. Therefore, {\em we define} the metric for
the tube dislocation as
\begin{equation}                                                  \label{eindum}
  ds^2=(1-v)^2dr^2+(r-u)^2d\vf^2+dz^2
\end{equation}
with the volume element
\begin{equation*}
  \vol=(1-v)(r-u),~~~~\text{where}~~g=\det g_{\mu\nu}.
\end{equation*}
Metric (\ref{eindum}) differs from the formal substitution of $y=r-u(r)$ in the
Euclidean metric $ds^2=dy^2+y^2d\vf^2+dz^2$ by the square of the $\dl$-function
in the $g_{rr}$ component. This procedure is a must in the geometric theory of
defects, because otherwise the Burgers vector can not be expressed as the
surface integral \cite{Katana05}. So, the metric component $g_{rr}(r)=(1-v)^2$
of tube dislocation is a continuous function, and the angular component
$g_{\vf\vf}=(r-u)^2$ has the jump across the cut.

Thus we solved linear elasticity field equations (\ref{eqsepu}) with boundary
conditions (\ref{ebotud}) describing the tube dislocation. The displacement
vector field (\ref{edifun}) and, consequently, induced metric (\ref{eindum})
do not depend on the elastic properties of media characterized by a Poisson
ratio. This demonstrates the universality of the problem.

Now we calculate geometric quantities for the tube dislocation.
The components of the metric are not differentiable functions, and hence the
calculation of geometric quantities involving derivatives and multiplications
is an ambiguous procedure. Therefore, we perform all calculations as if the
components were sufficiently smooth functions and see that all ambiguous terms
safely cancel in the final answer. It means that whatever regularization of the
components is applied the final answer does not depend on it.

Now we calculate Christoffel's symbols:
\begin{equation*}
  \widetilde\G_{\mu\nu\rho}=\widetilde\G_{\mu\nu}{}^\s g_{\s\rho}
  =\frac12(\pl_\mu g_{\nu\rho}+\pl_\nu g_{\mu\rho}-\pl_\rho g_{\mu\nu}).
\end{equation*}
Only four components of Christoffel's symbols differ from zero,
\begin{equation*}
\begin{split}
  \widetilde\G_{rrr}&=-v'(1-v),
\\
  \widetilde\G_{r\vf\vf}=\widetilde\G_{\vf r\vf}&=~~[1-v+l\dl(r-r_*)](r-u),
\\
  \widetilde\G_{\vf\vf r}&=-[1-v+l\dl(r-r_*)](r-u),
\end{split}
\end{equation*}
where we used relation (\ref{ereluv}). The nonzero components of Christoffel's
symbols with one raised index are as follows
\begin{equation*}
\begin{split}
  \widetilde\G_{rr}{}^r&=-\frac{v'}{1-v},
\\
  \widetilde\G_{r\vf}{}^\vf=\widetilde\G_{\vf r}{}^\vf&
  =~~\frac{1-v+l\dl(r-r_*)}{r-u},
\\
  \widetilde\G_{\vf\vf}{}^r&=-\frac{[1-v+l\dl(r-r_*)](r-u)}{(1-v)^2}.
\end{split}
\end{equation*}
Note that if the $\dl$-function was not dropped in $g_{rr}$ component
(\ref{eindum}), then we would have to divide on it, which is a forbidden
procedure.

The curvature tensor components in our notations are
\begin{equation*}
  \widetilde R_{\mu\nu\rho\s}=\pl_\mu\widetilde\G_{\nu\rho\s}
  -\pl_\nu\widetilde\G_{\mu\rho\s}
  +\widetilde\G_{\mu\rho}{}^\lm\widetilde\G_{\nu\s\lm}
  -\widetilde\G_{\nu\rho}{}^\lm\widetilde\G_{\mu\s\lm}.
\end{equation*}
There is only one nonvanishing independent component
\begin{equation}                                                  \label{ecurvt}
\begin{split}
  \widetilde R_{r\vf r\vf}&=\pl_r\widetilde\G_{\vf r\vf}
  +\widetilde\G_{rr}{}^r\widetilde\G_{\vf\vf r}
  -\widetilde\G_{\vf r}{}^\vf\widetilde\G_{r\vf\vf}=
\\
  &=l(r-u)\left[\dl'(r-r_*)+\frac{v'}{1-v}\dl(r-r_*)\right],
\end{split}
\end{equation}
where
\begin{equation*}
  \dl'(r-r_*)=\pl_r\dl(r-r_*)
\end{equation*}
is the derivative of the $\dl$-function.
The ambiguous terms with the squares $\dl^2(r-r_*)$ safely cancel.

It is interesting and quite important that the prescription to drop the
$\dl$-function from the $g_{rr}$ component in the metric for a tube dislocation
(\ref{eindum}) which comes from physical considerations \cite{Katana05}
determines the right way for mathematical treatment of the calculation of the
curvature. Note also that if the $\dl$-function was not dropped then the
curvature tensor would be identically zero because any diffeomorphism of the
Euclidean space $\MR^3$ leaves the curvature tensor equal to zero.

We see that curvature (\ref{ecurvt}) is zero everywhere except the cutting
surface, as it must be both from physical and mathematical standpoints.

The component of the curvature tensor (\ref{ecurvt}) is still an ambiguous
functions because the coefficient in front of $\dl'$ and $\dl$-functions are
not continuous.

The Ricci tensor has two nonvanishing components,
\begin{equation*}
\begin{split}
  \widetilde R_{rr}&=\frac l{r-u}\left[\dl'(r-r_*)+
  \frac{v'}{1-v}\dl(r-r_*)\right],
\\
  \widetilde R_{\vf\vf}&=\frac{l(r-u)}{(1-v)^2}
  \left[\dl'(r-r_*)+\frac{v'}{1-v}\dl(r-r_*)\right].
\end{split}
\end{equation*}
The scalar curvature is
\begin{equation*}
  \widetilde R=\frac{2l}{(r-u)(1-v)^2}
  \left[\dl'(r-r_*)+\frac{v'}{1-v}\dl(r-r_*)\right].
\end{equation*}
Einstein's equations,
\begin{equation}                                                  \label{enmode}
  \vol\left(\widetilde R_{\mu\nu}-\frac12g_{\mu\nu}\widetilde R\right)
  =-\frac12T_{\mu\nu},
\end{equation}
are identically satisfied except the $zz$ component,
\begin{equation}                                                  \label{enmotu}
  2l\left(\frac 1{1-v}\dl'(r-r_*)+\frac{v'}{(1-v)^2}\dl(r-r_*)\right)
  =T_{zz},
\end{equation}
where $T_{zz}$ is the source of the tube dislocation (the analog of the
energy-momentum tensor density in general relativity). Using the identity for an
arbitrary differentiable function $f\in\CC^1(\MR_+)$ on the positive real line,
\begin{equation*}
  f(r)\dl'(r-r_*)=f(r_*)\dl'(r-r_*)-\pl_r f\dl(r-r_*),
\end{equation*}                                                   \label{etzzco}
the singular part in the geometric expressions can be rewritten as
\begin{equation}                                                  \label{etusof}
  T_{zz}=\frac{2l}{1-v(r_*)}\dl'(r-r_*)=\frac{4lr_*}{2r_*-l}\dl'(r-r_*).
\end{equation}
We see that all ambiguous terms cancel ! In general, if components of a metric
are not continuous functions, then the curvature tensor components will be not
unambiguously defined, because they will have squares of the type $\dl^2$ and
products of the step function with $\dl$-function. Therefore, cancelation of all 
ambiguous terms in a geometric quantity is a big surprise. In the considered 
case, Christoffel's symbols and components of the curvature tensor are not well 
defined but the energy-momentum tensor is unambiguous. The same situation 
happens for a distribution of wedge dislocations with the same factor $\sqrt{g}$ 
in the definition of the energy-momentum tensor density (\ref{enmode}) 
\cite{Katana05}. The factor $\vol$ appears in Einstein's equations 
(\ref{enmode}) because $\dl$-function is not a function but a scalar density 
with respect to coordinate transformations.
\section{Summary of the geometric theory of defects}
In this section we briefly formulate the geometric theory of defects
developed in \cite{KatVol92,KatVol99,Katana03,Katana04} (for review see
\cite{Katana05}). This model treats defects in elastic media with a spin
structure entirely within differential geometry. At the moment it is developed
only for a static distribution of single defects as well as continuous
distribution of dislocations and disclinations, the basic equations coinciding
with equations of three-dimensional Euclidean gravity with torsion.

In the geometric theory of defects we assume that elastic media is a
three-dimensional manifold $\MM$ with a given Riemann--Cartan geometry. For
simplicity, we consider topologically trivial manifold $\MM\approx\MR^3$ which
is covered by an arbitrary curvilinear coordinate system $x^\mu$, $\mu=1,2,3$.
The basic variables are the triad field $e_\mu{}^i(x)$ and $\MS\MO(3)$-connection
$\om_\mu{}^{ij}(x)=-\om_\mu{}^{ji}(x)$ (Cartan variables). Note that we do not
have the displacement vector field $u^i(x)$ as an independent variable.
There is only geometry and nothing else in the geometric approach. The
torsion and curvature tensors are
\begin{align}                                           \label{ecurcv}
  T_{\mu\nu}{}^i&=\pl_\mu e_\nu{}^i-e_\mu{}^j\om_{\nu j}{}^i
                    -(\mu\leftrightarrow\nu),
\\                                                      \label{ecucav}
  R_{\mu\nu j}{}^i&=\pl_\mu \om_{\nu j}{}^i-\om_{\mu j}{}^k
                       \om_{\nu k}{}^i-(\mu\leftrightarrow\nu).
\end{align}
These geometric notions are $2$-forms and have straightforward physical
interpretation as the surface density of the Burgers and Frank vectors,
respectively, characterizing distributions of dislocations and disclinations.
If geometry is trivial, $T_{\mu\nu}{}^i=0$ and $R_{\mu\nu}{}^{ij}=0$, then there
are no defects in media. The nonzero torsion and curvature is the criteria for
the presence of dislocations and disclinations, respectively.

For a given Riemann--Cartan geometry $(\MM,e_\mu{}^i,\om_\mu{}^{ij})$, we have,
in fact, two connections on a manifold because, having the triad field at hand,
we can compute the metric $g_{\mu\nu}=e_\mu{}^i e_\nu{}^j\dl_{ij}$, where
$\dl_{ij}=\diag(+++)$ is the Euclidean metric, and Christoffel's symbols which
define the second connection on $\MM$. The corresponding second
$\MS\MO(3)$-connection can be expressed in terms of the triad field alone
\begin{equation}                                                  \label{ezertc}
  \tilde\om_{ijk}=\frac12(c_{ijk}-c_{jki}+c_{kij}),
\end{equation}
where
\begin{equation*}
  c_{\mu\nu}{}^i=-c_{\nu\mu}{}^i=-\pl_\mu e_\nu{}^i+\pl_\nu e_\mu{}^i
\end{equation*}
are the anholonomicity coefficients and transformations of Latin indices into
Greek ones are performed using the triad field and its inverse,
$c_{ijk}=e^\mu{}_i e^\nu{}_j c_{\mu\nu i}$. The anholonomicity coefficients
define the commutator of the orthonormal basis vector fields
$e_i=e^\mu{}_i\pl_\mu$,
\begin{equation*}
  [e_i,e_j]=c_{ij}{}^k e_k.
\end{equation*}
The second $\MS\MO(3)$-connection, $\tilde\om_\mu{}^{ij}$, differs from the
original connection, $\om_\mu{}^{ij}$, and therefore is denoted by the tilde
sign. It corresponds to zero torsion and is the unique solution of the equation
$T_{\mu\nu}{}^i=0$ with respect to connection.

This duality is well known in general relativity. Usually, we solve Einstein
equations with respect to metric $g_{\mu\nu}$, compute Christoffel's symbols
$\widetilde\G_{\mu\nu}{}^\rho$, curvature tensor $\widetilde R_{\mu\nu\rho\s}$,
and interpret gravity as nontrivial curvature of the space-time. The second
alternative is to solve Einstein's equations with respect to the tetrad field
$e_\mu{}^i$, assume that the curvature tensor of the space-time is identically
zero $R_{\mu\nu}{}^{ij}=0$ (the space of absolute parallelism), put the Lorentz
$\MS\MO(1,3)$-connection to zero (this fixes the invariance with respect to
local Lorentz rotations), and compute the torsion tensor
\begin{equation}                                                  \label{etocoa}
  T_{\mu\nu}{}^i=-c_{\mu\nu}{}^i=\pl_\mu e_\nu{}^i-\pl_\nu e_\nu{}^i.
\end{equation}
Hence, gravity is interpreted as nontrivial torsion in a space with absolute
parallelism. The invariance with respect to local Lorentz rotations can be
restored assuming that the Lorentz connection is not zero but a pure gauge.
The metric tensor is the same in both cases, and we always have two different
geometric interpretations of any solution of Einstein's equations.

Now we proceed with the geometric theory of defects. The equations of
equilibrium for dislocations and disclinations follow from the expression for
the free energy functional which was proposed in \cite{KatVol92}
\begin{equation}                                                  \label{eacfre}
  S=\int d^3x\left[\vol(-\kappa\widetilde R+2\g R^\Sa{}_{\mu\nu}R^{\Sa\mu\nu})
  -\frac12g_{\mu\nu}T^{\mu\nu}\right],
\end{equation}
where the first term is the Hilbert--Einstein action depending only on the triad
field, $R^\Sa{}_{\mu\nu}=\frac12(R_{\mu\nu}-R_{\nu\mu})$ is the antisymmetric
part of the Ricci tensor which is nonzero for nontrivial torsion, and
$T^{\mu\nu}$ is the source for dislocations (the energy-momentum tensor density
in general relativity). $\kappa$ and $\g$ are two coupling constants (which are
not determined yet, but do not alter the following calculations). If needed, the
Hilbert--Einstein term in the action can be rewritten using the identity
\begin{equation}                                                  \label{eidcur}
  R(e,\om)+\frac14T_{ijk}T^{ijk}-\frac12T_{ijk}T^{kij}
  -T_iT^i-\frac2{\vol}\pl_\mu(\vol T^\mu)=\widetilde R(e),~~~~g=\det g_{\mu\nu},
\end{equation}
where $T_i=T_{ji}{}^j$, in terms of the the scalar curvature $R(e,\om)$ which 
depends on the triad field and $\MS\MO(3)$-connection and torsion squared terms.

The equations of equilibrium following from the expression for a free energy
(\ref{eacfre}) are covariant and of second order for $e_\mu{}^i$ and
$\om_\mu{}^{ij}$. They define the triad field and $\MS\MO(3)$-connection up
to general coordinate transformations and local rotations. To fix the solution
uniquely, we must impose gauge conditions. In the geometric theory of defects we
impose the elastic gauge to fix diffeomorphisms and the Lorentz gauge for local
rotations. To write these conditions as equations, we denote the triad, metric,
Christoffel's symbols and $\MS\MO(3)$-connection for the flat Euclidean space
$\MR^3$ in an arbitrary curvilinear but fixed coordinate system by
$\overset\circ e_\mu{}^i$, $\overset\circ g_{\mu\nu}$,
$\overset\circ\G_{\mu\nu}{}^\rho$, and $\overset\circ\om_\mu{}^{ij}$,
respectively. Then the gauge conditions are first order differential equations
\cite{Katana03,Katana04},
\begin{align}                                                     \label{efouad}
  \overset\circ g{}^{\mu\nu}\overset{\circ}{\nb}_\mu e_{\nu i}
  +\frac\s{1-2\s}\overset\circ e{}^\mu{}_i\overset\circ\nb_\mu e^T&=0,
\\                                                                \label{elorog}
  \overset\circ g{}^{\mu\nu}\overset{\circ}{\nb}_\mu\om_\nu{}^{ij}&=0,
\end{align}
where the covariant derivative $\overset\circ\nb_\mu$ includes both
Christoffel's symbols $\overset\circ\G_{\mu\nu}{}^\rho$ and
$\MS\MO(3)$-connection $\overset\circ\om_\mu{}^{ij}$ acting on Greek and Latin
indices, and $e^T=\overset\circ e{}^\mu{}_ie_\mu{}^i$. These conditions are
invariant with respect to coordinate transformations in the Euclidean space
$\MR^3$ but fix the solution of the equilibrium equations in these coordinates.
For example, we can use cylindrical or spherical coordinates in $\MR^3$
depending on the problem. For a fixed coordinate system, the gauge conditions
(\ref{efouad}) and (\ref{elorog}) almost uniquely define the solution of
covariant equations of equilibrium following from the action (\ref{eacfre}). In
fact, some additional assumptions depending on the problem are needed for the
unique definition, because gauge conditions themselves are differential
equations. The examples which are considered later clarify the situation.

The gauge conditions (\ref{efouad}), (\ref{elorog}) have straightforward
physical meaning \cite{Katana03,Katana04} and play a crucial role in the
geometric theory of defects. Suppose that defects are absent, that is, curvature
and torsion are equal to zero: $R_{\mu\nu}{}^{ij}=0$ and $T_{\mu\nu}{}^i=0$.
Then the $\MS\MO(3)$-connection is locally a pure gauge
\begin{equation*}
  \om_{\mu j}{}^i=\pl_\mu S^{-1}{}_j{}^k S_k{}^i,~~~~S_j{}^i(x)\in\MS\MO(3),
\end{equation*}
and there is a vector field $y^i(x)$ such that
\begin{equation*}
  e_\mu{}^i=\pl_\mu y^j S_j{}^i.
\end{equation*}
The functions $y^i(x)$ describe a transformation to Cartesian coordinates in
the Euclidean space $\MR^3$. The orthogonal matrix $S_j{}^i$ can be
parameterized by the rotational angle field $\om^{ij}(x)\in\Gs\Go(3)$ describing
the spin structure of media. In those domains of media where curvature and
torsion vanish the media is described by the vector field $y^i(x)$ and the
rotational angle field $\om^{ij}$. The equations of equilibrium following from
the free energy expression (\ref{eacfre}) are identically satisfied
\cite{KatVol92}, and we are left only with the gauge conditions (\ref{efouad}),
(\ref{elorog}) for the vector field $y^i$ and rotational angle field $\om^{ij}$.
Then the Lorentz gauge (\ref{elorog}) for the $\om^{ij}$ reduces to the
principal chiral $\MS\MO(3)$-field model describing the spin structure of media
\cite{Katana04}. In the linear approximation with respect to the displacement
vector field $u^i$ defined in (\ref{eeldef}), the triad field in Cartesian
coordinates can be chosen as
\begin{equation*}
  e_{\mu i}\simeq \dl_{\mu i}-\frac12(\pl_\mu u_i+\pl_i u_\mu),
\end{equation*}
and the elastic gauge (\ref{efouad}) coincides with equations (\ref{eqsepu}) of
linear elasticity theory.  Equations (\ref{efouad}) themselves are infinite
power series in the displacement vector field $u^i$ of nonlinear elasticity
theory. Thus we chose two theories, nonlinear elasticity theory and principal
chiral $\MS\MO(3)$-field model as the gauge conditions in the geometric theory
of defects.

This is in contrast with the main idea of general relativity which is to
consider all coordinate systems as equivalent. In the geometric theory of
defects, we have a preferred coordinate system defined by the elastic
(\ref{efouad}) and Lorentz (\ref{elorog}) gauge conditions. The advantage of
this approach is that the theory automatically contains all results obtained
within the nonlinear elasticity theory for elastic deformations and principal
chiral $\MS\MO(3)$-field model for the spin structure.

There is another important point. At present, there is no commonly acknowledged
fundamental theory for continuous distribution of defects. A distribution of
single dislocations can be described using the notion of the displacement vector
field within the elasticity theory. This approach fails for description of
continuous distribution of defects because the notion of the displacement vector
field disappears and there is no reason to talk about any theory at all. The
same happens with disclinations. Indeed, for a distribution of single
disclinations, the rotational angle field exists almost everywhere and we can
use principal chiral $\MS\MO(3)$-field model for the spin structure (or any
other of the existing models). If disclinations are distributed continuously
then the rotational angle field does not exist. Therefore, it is not
possible to define a model for continuous distribution of disclinations using
the rotational angle field as a variable.

In the geometric theory of defects which pretends to be a fundamental one, we
start from the opposite point. We drop the notions of the displacement vector
$u^i$ and rotational angle $\om^{ij}$ fields as the basic variables of the
theory. Instead, we choose the triad field $e_\mu{}^i$ and
$\MS\MO(3)$-connection $\om_\mu{}^{ij}$ as the only independent variables and
postulate covariant equations for them following from the variation of the
action (\ref{eacfre}). These equations define a solution up to general
coordinate transformations and local rotations. We choose two models to fix the
solution uniquely: the nonlinear elasticity theory for elastic deformations and
the principal chiral $\MS\MO(3)$-field model for the spin structure. And this is
exactly what is needed. Indeed, the purpose of elasticity theory is to find the
displacement vector field. This field parameterizes diffeomorphisms and
therefore fixes the coordinate system. Similarly, the principal chiral
$\MS\MO(3)$-field model defines the rotational angle field and hence fixes local
rotations. Though the models are present in the geometric theory of defects,
they are rewritten in terms of the triad field and $\MS\MO(3)$-connection, and
the notions of the displacement vector and rotational angle fields are still
absent. For continuous distribution of defects, these fields can not be defined
at all because the curvature and torsion differ from zero. The displacement
vector and rotational angle fields can be defined only in those domains where
curvature and torsion are zero, i.e., defects are absent. If so, the equations
of equilibrium following from the free energy (\ref{eacfre}) are identically
satisfied, and we are left with the old highly appreciated elasticity theory for
elastic deformations \cite{Katana03} and principal chiral $\MS\MO(3)$-field
model for the spin structure \cite{Katana04}.
\section{Tube dislocation in the geometric theory of defects}     \label{stugeo}
In this section we apply the geometric theory of defects for the description of
tube dislocations. To this end we must solve equations of equilibrium following
from the action (\ref{eacfre}) in the coordinate system defined by the elastic
(\ref{efouad}) and Lorentz (\ref{elorog}) gauge conditions. The tube dislocation
is determined by the source $T_{\mu\nu}$ which will be specified later on.

The tube dislocation problem corresponds to elastic media without a spin
structure and hence without disclinations. In this case, the curvature tensor is
zero $R_{\mu\nu}{}^{ij}=0$ (but not the $\widetilde R_{\mu\nu}{}^{ij}$ !), the
equations of equilibrium obtained from (\ref{eacfre}) by variation with respect
to $\om_\mu{}^{ij}$ are identically satisfied, and we can safely put
$R^\Sa{}_{\mu\nu}=0$ in the action (\ref{eacfre}). Afterwards, we vary the action
with respect to the triad field $e_\mu{}^i$ and obtain Einstein's equations
(\ref{enmode}) if we put $\kappa=1$ for simplicity.

In the case of the tube dislocation, the metric has two Killing vector fields
$\pl_z$ and $\pl_\vf$ in cylindrical coordinates which correspond to
translations along $z$ axis and rotations in the $x,y$ plane. For solution of
Einstein's equations we choose the metric in diagonal form
\begin{equation}                                                  \label{eunmeg}
  g_{\mu\nu}=\begin{pmatrix}A^2 & 0 & 0 \\ 0 & B^2 & 0 \\ 0 & 0 & 1\end{pmatrix},
\end{equation}
where $A(r)$ and $B(r)$ are unknown positive functions of radius. This is
not the most general form of the metric which is consistent with the symmetry of
the problem but it is sufficient for our purpose. The corresponding triad field
can be also chosen in diagonal form with three nonzero components
\begin{equation}                                                  \label{etranz}
  e_r{}^{\hat r}=A,~~~~e_\vf{}^{\hat\vf}=B,~~~~e_z{}^{\hat z}=1,
\end{equation}
where indices are denoted as $\lbrace\mu\rbrace=(r,\vf,z)$ and
$\lbrace i\rbrace=(\hat r,\hat\vf,\hat z)$. The volume measure is
\begin{equation*}
  \sqrt g=AB.
\end{equation*}
If we did not impose the elastic gauge (\ref{efouad}) then the remaining freedom
in choosing the radial coordinate could be used to make further simplifications.
For example, we could choose $A=1$.

The following calculations are performed with $A$ and $B$ as differentiable
functions, and the obtained discontinuous solution will be justified by the
cancelation of all ambiguous terms in the equation for $A$ and $B$.

Christoffel's symbols have four nontrivial components,
\begin{equation*}
\begin{split}
  \widetilde\G_{rr}{}^r&=\frac{A'}A,
\\
  \widetilde\G_{r\vf}{}^\vf=\widetilde\G_{\vf r}{}^\vf&=\frac{B'}B,
\\
  \widetilde\G_{\vf\vf}{}^r&=-\frac{BB'}{A^2}.
\end{split}
\end{equation*}
The curvature tensor has only one independent nontrivial component
\begin{equation*}
  \widetilde R_{r\vf r\vf}=BB''-\frac{A'BB'}A.
\end{equation*}
The nonzero Ricci tensor components and scalar curvature are
\begin{align*}
  \widetilde R_{rr}&=\frac{B''}B-\frac{A'B'}{AB},
\\
  \widetilde R_{\vf\vf}&=\frac{BB''}{A^2}-\frac{A'BB'}{A^3},
\\
  \widetilde R&=\frac2{AB}\left(\frac{B''}A-\frac{A'B'}{A^2}\right).
\end{align*}

We suppose that the source for the tube dislocation has only one nonzero
component
\begin{equation}                                                  \label{etusou}
  T_{zz}=L\dl'(r-r_*),
\end{equation}
where $L$ and $r_*$ are two constants which characterize the strength and
position of a tube dislocation. This form of the source is prompted by the
elasticity theory (\ref{etusof}). In general, you can take any function you
want as the source of dislocations and obtain the corresponding triad field and
metric. This will be a solution of a different problem describing some
distribution (that may be continuous) of parallel dislocations with circular
symmetry.

The $rr$, $\vf\vf$, and nondiagonal components of Einstein's equations
(\ref{enmode}) for metric (\ref{eunmeg}) are identically satisfied, and the $zz$
component reduces to one ordinary differential equation
\begin{equation*}
  \left(\frac{B'}A\right)'=\frac12L\dl'(r-r_*).
\end{equation*}
This is a {\em linear} inhomogeneous equation with respect to the combination of
diagonal metric components $B'/A$. Therefore, the consideration of singular
source (\ref{etusou}) is justified. This equation is well defined if $A(r)$ is a
positive and continuous function and has a general solution
\begin{equation}                                                  \label{efiibn}
  B'=\frac12A(r_*)L\dl(r-r_*)+c_1A,~~~~c_1=\const,
\end{equation}
where $c_1$ is the integration constant. The obtained expression can be further
integrated
\begin{equation}                                                  \label{esolba}
  B=\frac12A(r_*)L\theta(r-r_*)+c_1\int_0^r ds A(s)+c_2,~~~~c_2=\const,
\end{equation}
where $\theta$ is the step function
\begin{equation}                                                  \label{estepf}
  \theta(r-r_*)=\begin{cases} 0, & r\le r_*, \\ 1, & r>r_*. \end{cases}
\end{equation}
Thus we solved Einstein's equations for a tube dislocation. This solution is
determined up to one arbitrary positive and continuous function $A(r)$ and two
constants of integration $c_{1,2}$. An arbitrary function in the solution
reflects the remaining freedom in choosing the radial coordinate and has to be
fixed by the elastic gauge.

In essence, Einstein's equations for the source (\ref{etusou}) determine only
the jump in the $e_\vf{}^{\hat\vf}$ component of the triad. The arbitrary
function $e_r{}^{\hat r}=A(r)$ describes the freedom to choose the radial
coordinate which is still left.

Now we impose the elastic gauge (\ref{efouad}). In cylindrical coordinates, the
flat triad field can be chosen in diagonal form with components
\begin{equation*}
  \overset\circ e_r{}^{\hat r}=1,~~~~\overset\circ e_\vf{}^{\hat\vf}=r,
  ~~~~\overset\circ e_z{}^{\hat z}=1.
\end{equation*}
It defines flat Christoffel's symbols $\overset\circ\G_{\mu\nu}{}^\rho$ and
$\MS\MO(3)$-connection $\overset\circ\om_{\mu i}{}^j$ with the following nonzero
components:
\begin{equation*}
\begin{split}
  \overset\circ\G_{r\vf}{}^\vf&=\overset\circ\G_{\vf r}{}^\vf=\frac1r,~~~~~~
  \overset\circ\G_{\vf\vf}{}^r=-r,
\\
  \overset\circ\om_{\vf\hat r}{}^{\hat\vf}&
  =-\overset\circ\om_{\vf\hat\vf}{}^{\hat r}=1.
\end{split}
\end{equation*}
Substitution of vielbein (\ref{etranz}) into gauge condition (\ref{efouad})
yields the Euler differential equation for $A(r)$
\begin{equation}                                                  \label{eulbou}
  A'+\frac Ar-\frac B{r^2}
  +\frac\s{1-2\s}\left(A'+\frac{B'}r-\frac B{r^2}\right)=0,
\end{equation}
where $B(r)$ is given by (\ref{esolba}).

We are looking for classical solutions of this equation inside the cutting
surface, $A_\text{in}$, and outside it, $A_\text{ex}$, with the
``asymptotically flat'' boundary conditions:
\begin{equation}                                                  \label{easfbo}
  B_\text{in}|_{r=0}=0,~~~~0<A_\text{in}|_{r=0}<\infty,
  ~~~~B_\text{ex}|_{r\to\infty}=r.
\end{equation}
The boundary conditions at $r=0$ are the same as for the Euclidean metric.
Moreover, on the cutting surface we impose the boundary conditions:
\begin{equation}                                                  \label{ecutbo}
  A_\text{in}|_{r=r_*}=A_\text{ex}|_{r=r_*},~~~~
  B_\text{in}|_{r=r_*}+\frac{LA(r_*)}2=B_\text{ex}|_{r=r_*}.
\end{equation}
The first matching condition provides the equality of the normal elastic forces. 
The second matching condition is the consequence of Eq.(\ref{esolba}) and 
provides the jump of the $e_\vf{}^{\hat\vf}$ component of the triad. Boundary 
conditions (\ref{easfbo}), (\ref{ecutbo}) are analogous to boundary conditions 
(\ref{ebotud}) in the elasticity theory problem.

Substitution of $B$ from Eq.(\ref{esolba}) into the first boundary condition
(\ref{easfbo}) yields $c_2=0$.

Eq.(\ref{eulbou}) is more easily solved with respect to function $B$ instead of
$A$. Inside and outside the cutting surface, $B'=c_1A$ as the consequence of
(\ref{efiibn}). Then Eq.(\ref{eulbou}) reduces to
\begin{equation}                                                  \label{ebfute}
  \frac{B''}{c_1}+\frac{B'}{c_1r}-\frac B{r^2}+\frac\s{1-2\s}
  \left(\frac{B''}{c_1}+\frac{B'}r-\frac B{r^2}\right)=0.
\end{equation}
This equation coincides with the equation for the wedge dislocation
\cite{Katana05}, the difference $c_1-1$ playing the role of the deficit angle
of the conical singularity. It has a general solution
\begin{equation}                                                  \label{egeuls}
  B=D_1r^{\g_1}+D_2r^{\g_2},~~~~D_{1,2}=\const,
\end{equation}
depending on two integration constants $D_{1,2}$ and where $\g_{1,2}$ are the
roots of the quadratic equation
\begin{equation*}
  \g^2+\frac{(c_1-1)\s}{1-\s}\g-c_1=0,
\end{equation*}
which has two real roots for $c_1>0$, with different signs: the positive root
$\g_1$ and the negative one $\g_2$.

In the internal region, we have $D_2=0$ and $\g_1=1$ as the consequence of the
first two conditions in (\ref{easfbo}), the condition $\g_1=1$ being equivalent
to $c_1=1$ for $\s\ne1/2$. Hence, the solution of Einstein's equations in the
internal region is
\begin{equation}                                                  \label{einsol}
  B_\text{in}=D_1r,~~~~A_\text{in}=D_1.
\end{equation}
It depends on one arbitrary constant $D_1$. So, the first two boundary
conditions in (\ref{easfbo}) fix two constants of integration of Einstein's
equations, $c_1=1$ and $c_2=0$, which are the same in both internal and external
regions and one constant of integration of elastic gauge, $D_2=0$.

To reduce the number of indices, we denote constants of integration in
Eq.(\ref{egeuls}) in the external region by new letters,
\begin{equation*}
  B_\text{ex}=E_1r^{\g_1}+E_2r^{\g_2},~~~~E_{1,2}=\const.
\end{equation*}
The constant of integration $c_1=1$ was already fixed in the internal region,
and therefore $\g_1=1$ and $\g_2=-1$. The third asymptotic condition
(\ref{easfbo}) determines $E_1=1$ and yields solution in the external region,
\begin{equation}                                                  \label{eousol}
  B_\text{ex}=r+\frac{E_2}r,~~~~A_\text{ex}=1-\frac{E_2}{r^2},
\end{equation}
which also depends on one arbitrary constant $E_2$.

To express constants $D_1$ and $E_2$ in terms of parameters $L$ and $r_*$
characterizing the source, we use matching conditions. The first condition in
Eqs.(\ref{ecutbo}) relates constants $D_1$ and $E_2$. Denoting $D_1=1-a$ and
$E_2=b$, we obtain Eq.(\ref{eglsur}). Finally, the second matching condition in
Eq.(\ref{ecutbo}) yields
\begin{equation*}
  L=2\frac{B_\text{ex}(r_*)-B_\text{in}(r_*)}{A(r_*)}=\frac{4lr_*}{2r_*-l},
\end{equation*}
where $l=2\sqrt{ab}$, coincides with the coefficient in front of the
$\dl$-function in source term (\ref{etusou}).

Thus, we solved the problem for the tube dislocation within the geometric theory
of defects. The  triad field is given by (\ref{etranz}) where functions $A$ and
$B$ are given by Eqs.(\ref{einsol}) and (\ref{eousol}). The induced metric
(\ref{eunmeg}) coincides exactly with the induced metric obtained within the
elasticity theory (\ref{eindum}). It is important that we obtain the induced
metric by solving Einstein's equations in the elastic gauge knowing nothing
about the displacement vector field. If needed, we can reconstruct the
displacement vector field in the internal and external regions by solving
equation
\begin{equation*}
  \frac{\pl y^i}{\pl x^\mu}=e_\mu{}^i
\end{equation*}
with appropriate boundary conditions, where $y^i$ is defined in (\ref{eeldef}).
The solution of this equation exists, because the curvature tensor is zero in
both regions, $\widetilde R_{\mu\nu\rho\s}=0$. In terms of metric, we have to
find such coordinate systems in the internal and external regions were metric
becomes Euclidean. This can be easily done. The essential two dimensional part
of the metric in the inside and outside regions of the gluing surface are
\begin{align*}
  dl_\text{in}^2&=(1-a)^2dr^2+(1-a)^2r^2d\vf^2,
\\
  dl_\text{ex}^2&=\left(1-\frac br\right)^2dr^2+\left(r+\frac br\right)^2d\vf^2.
\end{align*}
Introducing new coordinate $y=(1-a)r$ in the internal and $y=r+b/r$ in the
external regions, the metric is brought to the Euclidean form
$dl^2=dy^2+y^2d\vf^2$ in both regions. Afterwards, we immediately obtain the
displacement vector field (\ref{edifun}) using definition (\ref{eeldef}).

So, we solved the problem for the tube dislocation within the elasticity theory
and geometric theory of defects. The result is the same in both approaches,
although this coincidence is not an automatic rule. For example, the induced
metric for the wedge dislocation obtained in the geometric theory of defects
\cite{Katana03} differs essentially from that in the elasticity theory and
reproduces the elasticity theory result only in the linear approximation. This
happens because the elastic gauge conditions (\ref{efouad}) coincide with
equations of the nonlinear elasticity theory for the displacement vector field.

There is a natural question: why should we use the sophisticated geometric
theory of defects if the ordinary elasticity theory works ? The answer is the
following. The elasticity theory works quite well for a single defect or in the
case of several single defects. If there are many single defects than the
boundary conditions become so complicated that there is no hope to solve the
corresponding problem. For example, we do not know the solution for arbitrary
distribution of parallel wedge dislocations within the elasticity theory whereas
this problem has a simple solution in the geometric theory of defects
\cite{KatVol92}. There is also a more important reason. Suppose that we have a
continuous distribution of defects, then the displacement vector field does not
exist and we can not even pose the problem within the elasticity theory. At
the same time, the problem for continuous distribution in the geometric theory
of defects is well posed: we have expression for the free energy (\ref{eacfre})
and gauge conditions (\ref{efouad}), (\ref{elorog}). The only difference is that
we have to consider a continuous source of defects $T_{\mu\nu}$ (the
energy-momentum tensor density).

The surface density of the Burgers vector is given by torsion components
(\ref{etocoa}). Simple calculations show that torsion for the tube dislocation
has only one independent nontrivial component,
\begin{equation}                                                  \label{etotud}
  T_{r\vf}{}^{\hat\vf}=-T_{\vf r}{}^{\hat\vf}=1-v+l\dl(r-r_*),
\end{equation}
which is singular on the cutting surface, but is nontrivial in the internal and
external regions. Projection of the total Burgers vector on the $x$ axis is
given by the integral and equals zero,
\begin{equation*}
  b^x=-\int_{\MR^2} \!\!dx^\mu\wedge dx^\nu T_{\mu\nu}{}^{\hat\vf}\sin\vf
  =\int_0^\infty\!\!\! dr\,r\int_0^{2\pi}\!\!\!d\vf\,T_{r\vf}{}^{\hat\vf}\sin\vf=0.
\end{equation*}
Similarly, its projection on any other straight line crossing the origin is
identically zero. Therefore the total Burgers vector is zero.
\section{Tube dislocation in General Relativity}
The main result of previous sections is the space metric (\ref{eindum})
describing a tube dislocation. We proved that it satisfies three-dimensional
Einstein's equations (\ref{enmode}) with the source (\ref{etusou}). The
generalization to four-dimensional general relativity is straightforward. We
assume that the tube dislocation does not move, i.e., it is static. The
corresponding metric is
\begin{equation}                                                  \label{etufou}
  ds^2=dt^2-(1-v)^2dr^2-(r-u)^2d\vf^2-dz^2,
\end{equation}
where the functions $v$ and $u$ are defined in (\ref{edefiu}) and (\ref{edifun}).
The component $g_{\vf\vf}$ is not continuous and $g_{rr}$ is continuous but not
differentiable. It is an easy exercise to check that all components of
Christoffel's symbols and curvature tensor having at least one time index are
identically zero for metric (\ref{etufou}). Therefore this metric satisfies
Einstein's equations
\begin{equation}                                                  \label{eeitub}
  \sqrt{|g|}\left(\widetilde R_{\al\bt}-\frac12g_{\al\bt}\widetilde R\right)
  =-\frac12T_{\al\bt}, ~~~~~~\al,\bt=0,1,2,3,
\end{equation}
where Greek letters from the beginning of the alphabet take values from 0 to 3.
The matter energy-momentum tensor density has only two nonvanishing components,
\begin{equation}                                                  \label{enmotu}
  T_{00}=-T_{zz}=L\dl'(r-r_*).
\end{equation}
It is a remarkable feature that all ambiguous terms in the left hand side of
Eq.(\ref{eeitub}) for the metric (\ref{etufou}) cancel. So, metric
(\ref{etufou}) satisfies Einstein's equations and its physical meaning is clear
from previous sections: it describes a static tube dislocation.
\section{Conical tube dislocation}
In previous sections, we described the tube dislocation in the framework of
elasticity theory and geometric theory of defects. The results turned out to be
identical. Here we describe another tube dislocations which are called conical
tube dislocations because they have asymptotic of conical singularity at
infinity. They are described entirely within the geometric theory of defects,
the solution of the problem in elasticity theory being unknown.

In this section, integration constants are denoted by the same letters as in
section \ref{stugeo} though they have different values.

Consider three dimensional Euclidean Einstein's equations (\ref{enmode}) with
the source term having only one nonzero component
\begin{equation}                                                  \label{esoutu}
  T_{zz}=2\Theta\dl(r-r_*),~~~~\Theta=\const,
\end{equation}
in cylindrical coordinates. In contrast to previous considerations, we changed
the derivative of the $\dl$-function in (\ref{etusou}) into the $\dl$-function
itself. We are looking for solutions of this problem which has translational and
circular symmetry. The constant $\Theta$ is interpreted as the deficit angle of
conical singularity corresponding to the asymptotic at $r\to\infty$.

This problem is equivalent to the problem of the static conical tube dislocation
in General Relativity with the energy-momentum tensor
\begin{equation}                                                  \label{enmoco}
  T_{00}=-T_{zz}=2\Theta\dl(r-r_*).
\end{equation}
Physical meaning of $2\Theta$ is the surface energy density of the thin cylinder
of radius $r_*$. For usual matter distribution, $\Theta>0$. However, we consider
both cases because mathematics does not depend much on the sign of $\Theta$.

We choose the metric in the diagonal form as in Eq.(\ref{eunmeg}). Then the
whole system of Einstein's equations reduce to one linear ordinary differential
equation
\begin{equation*}
  \left(\frac{B'}A\right)'=\Theta\dl(r-r_*).
\end{equation*}
It is easily integrated:
\begin{align}                                                     \label{efibpr}
  B'&=\Theta A\theta(r-r_*)+c_1A,
\\                                                                \label{esebpr}
  B&=\Theta\int_{r_*}^r dsA(s)+c_1\int_0^r dsA(s)+c_2,~~~~c_{1,2}=\const,
\end{align}
where $\theta(r-r_*)$ is the step function (\ref{estepf}) and $c_{1,2}$ are two
integration constants. So, Einstein's equations define the $e_\vf{}^{\hat\vf}$
component of the triad in terms of the $e_r{}^{\hat r}$ component, which can be
an arbitrary positive function. If $A(r)$ is a continuous function then $B(r)$
is also continuous due to Eq.(\ref{esebpr}), but its derivative has a jump
(\ref{efibpr}).

To specify the solution uniquely, we impose the elastic gauge (\ref{efouad})
which reduces to Eq.(\ref{eulbou}). We are looking for solutions of this
equation inside the cutting surface, $A_\text{in}$, and outside it,
$A_\text{ex}$. In elasticity theory, the wedge dislocation corresponds to an
infinite cylinder of finite radius, $0<r<R$, because stresses are divergent at
infinity \cite{Katana03}. We suppose that $R>r_*$. Therefore we impose the
following boundary conditions
\begin{equation}                                                  \label{ebocot}
  B_\text{in}|_{r=0}=0,~~~~0<A_\text{in}|_{r=0}<\infty,~~~~
  A_\text{ex}|_{r=R}=1.
\end{equation}
The last boundary condition implies the absence of external normal forces on the
boundary of the cylinder. We also impose matching conditions on the cutting
surface
\begin{equation}                                                  \label{emacos}
  A_\text{in}|_{r=r_*}=A_\text{ex}|_{r=r_*},~~~~
  B_\text{in}|_{r=r_*}=B_\text{ex}|_{r=r_*}.
\end{equation}

In the internal region, $B'=c_1A$ due to Eq.(\ref{efibpr}). Therefore the
elastic gauge reduces to Eq.(\ref{ebfute}) with the same boundary conditions as
for the tube dislocation considered earlier. Hence, $c_1=1$ and $c_2=0$ and the
solution in the internal region is given by Eq.(\ref{einsol}) as before.

In the external region, $B'=\al A$, where $\al=1+\Theta$, because $c_1=1$. A
general solution to Eq.(\ref{ebfute}) is
\begin{equation*}
  B_\text{ex}=E_1r^{\g_1}+E_2r^{\g_2},
\end{equation*}
where $\g_{1}>0$ and $\g_2<0$ are roots of the equation
\begin{equation}                                                  \label{equaga}
  \g^2+\frac{\Theta\s}{1-\s}\g-\al=0,~~~~\al=1+\Theta.
\end{equation}

The third boundary condition in Eq.(\ref{ebocot}),
\begin{equation}                                                  \label{ethbou}
  \frac{E_1\g_1}\al R^{\g_1-1}+\frac{E_2\g_2}\al R^{\g_2-1}=1,
\end{equation}
relates $E_1$ to $E_2$ for a given $R$.

Afterwards, constants $D_1$ and $E_2$ are determined through $\Theta$ and $r_*$
by matching conditions (\ref{emacos})
\begin{align}                                                     \label{ematoc}
  \frac{E_1\g_1}\al r_*^{\g_1-1}+\frac{E_2\g_2}\al r_*^{\g_2-1}&=D_1,
\\                                                                \label{ematsc}
  E_1r_*^{\g_1}+E_2r_*^{\g_2}&=D_1r_*.
\end{align}
In practice, we first solve Eqs.(\ref{ematoc}),(\ref{ematsc}):
\begin{equation}                                                  \label{esolec}
\begin{split}
  E_1&=~~D_1\frac{\al-\g_2}{\g_1-\g_2}r_*^{-\g_1+1},
\\
  E_2&=-D_1\frac{\al-\g_1}{\g_1-\g_2}r_*^{-\g_2+1}.
\end{split}
\end{equation}
The constant $D_1$ is found after substitution of these solutions into
Eq.(\ref{ethbou}),
\begin{equation}                                                  \label{esolfo}
  D_1=\frac\al{\g_1\frac{\al-\g_2}{\g_1-\g_2}\left(\frac R{r_*}\right)^{\g_1-1}
  -\g_2\frac{\al-\g_1}{\g_1-\g_2}\left(\frac R{r_*}\right)^{\g_2-1}}.
\end{equation}

So, the solution for a conical tube dislocation is
\begin{equation}                                                  \label{econtu}
  ds^2=A^2dr^2+B^2d\vf^2+dz^2,
\end{equation}
where
\begin{align}                                                     \label{esoame}
  A&=\begin{cases} D_1, & 0\le r\le r_*,\\
    \frac1\al(E_1\g_1r^{\g_1-1}+E_2\g_2r^{\g_2-1}), & r_*\le r\le R,
\end{cases}
\\                                                                \label{esobme}
    B&=\begin{cases} D_1r, & 0\le r\le r_*, \\
    E_1r^{\g_1}+E_2r^{\g_2},~~~~~~~~~~~~~~~& r_*\le r\le R, \end{cases}
\end{align}
where constants $D_1,E_{1,2}$ are given by Eqs.(\ref{esolec}), (\ref{esolfo})
in terms of constants $\Theta$, $r_*$, and $R$ defining the problem. The
components of the metric are continuous functions, but first derivative of $B$
has the jump $\Theta$ at $r=r_*$.

Now we give physical interpretation of the constant $\Theta$. Suppose that the
radius $R$ of the cylinder is large, $R\gg r_*$. Then near the boundary of the
cylinder $r\sim R$ the constants become
\begin{equation*}
\begin{split}
  D_1&\simeq\frac{\al(\g_1-\g_2)}{\g_1(\al-\g_2)}
  \left(\frac{r_*}R\right)^{\g_1-1},
\\
  E_1&\simeq\frac\al{\g_1R^{\g_1-1}},
\end{split}
\end{equation*}
and metric (\ref{econtu}) is asymptotically
\begin{equation}                                                  \label{easmet}
  ds^2=\left(\frac rR\right)^{2\g_1-2}
  \left(dr^2+\frac{\al^2}{\g_1^2}r^2d\vf^2\right)+dz^2.
\end{equation}
This is precisely the metric for the wedge dislocation corresponding to the
conical singularity with the deficit angle $\Theta$ \cite{Katana05}. Thus the
constant $\Theta$ standing in front of the $\dl$-function of the source
(\ref{esoutu}) coincides with the wedge which is removed, $-2\pi<\Theta<0$, or
added, $\Theta>0$, to the media.

From physical standpoint, we have found the metric for a conical tube
dislocation which defines elastic stresses around the defect. Note that this
solution does depend on the elastic properties of media in contrast to metric
(\ref{eindum}) obtained earlier. The metric can be considered as a solution to 
the problem even without knowing the displacement vector field distribution. 
However, an explicit construction of the displacement vector is very instructive 
and helps to imagine the defect creation (cut and paste procedure).

In the internal region the two-dimensional part of metric (\ref{econtu}) is
\begin{equation*}
  dl^2_\text{in}=D_1^2dr^2+D_1^2r^2d\vf^2.
\end{equation*}
In terms of new radial coordinate $y=D_1r$, the metric becomes Euclidean,
\begin{equation}                                                  \label{euctwm}
  dl^2_\text{in}=dy^2+y^2d\vf^2,
\end{equation}
with the gluing surface at $r=r_*$ corresponding to $r_1=y(r_*)=D_1r_*$. It
means that the cylinder $y\le r_1$ in the Euclidean space $y,\vf,z$ is mapped
into the internal region of the conical tube dislocation. The displacement
vector field (\ref{eeldef}) has only one nonzero component $u^r=r-y=(1-D_1)r$.

In the external region, the two dimensional part of the metric is
\begin{equation*}
  dl_\text{ex}=\frac1{\al^2}(E_1\g_1r^{\g_1-1}+E_2\g_2r^{\g_2-1})^2dr^2
  +(E_1r^{\g_1}+E_2r^{\g_2})^2d\vf^2.
\end{equation*}
Introducing new coordinates,
\begin{equation*}
  y=\frac1\al(E_1r^{\g_1}+E_2r^{\g_2}),~~~~\vf'=\al\vf,
\end{equation*}
the metric is brought to the Euclidean form
\begin{equation*}
  dl^2_\text{ex}=dy^2+y^2d\vf^{\prime 2}.
\end{equation*}
The range of new coordinates is
\begin{equation*}
  r_2\le y<\infty,~~~~0\le\vf'<2\pi\al,
\end{equation*}
where
\begin{equation*}
  r_2=\frac1\al(E_1r_*^{\g_1}+E_2r_*^{\g_2}).
\end{equation*}
It means that the external part of the cylinder $y>r_2$ of the Euclidean space
without the wedge $2\pi\al<\vf'<2\pi$ is mapped into the external region of the
conical tube dislocation. In this region, the displacement vector field has two
nonvanishing components,
\begin{equation*}
  u^r=r-\frac1\al(E_1r^{\g_1}+E_2r^{\g_2}),~~~~u^\vf=-\Theta\vf.
\end{equation*}

It is easily checked that circumference of the internal circle is equal to the
remained part of the external circle, $2\pi r_1=2\pi r_2\al$, in accordance with
Eq.(\ref{ematsc}). This provides the continuity of the $g_{\vf\vf}$ component of
the metric on the gluing surface.

\begin{figure}[h,b,t]
\hfill\includegraphics[width=.35\textwidth]{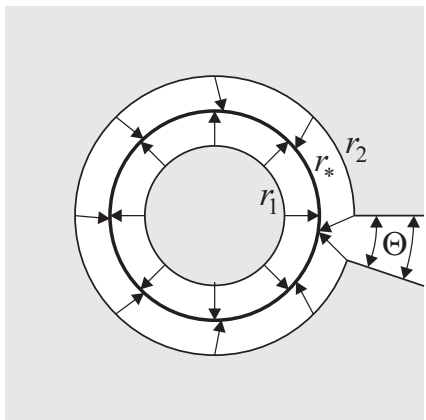}
\hfill {}
\\
\centering \caption{Conical tube dislocation for negative deficit angle
$\Theta$.}
\label{fcontu}
\end{figure}

The process of conical defect tube creation is shown in Fig.\ref{fcontu}. For
negative deficit angle $\Theta$, we take the Euclidean space $\MR^3$ with
cylindrical coordinates $y,\vf',z$, cut out the thick cylinder $r_1<y<r_2$ and
the wedge $2\pi\al<\vf'<2\pi$, and glue the sides of the cuts symmetrically as
shown in the figure. Afterwards, the media comes to some equilibrium state
governed by the elastic gauge which is called the conical tube dislocation. For
positive deficit angle, the wedge is added to the external part of the media,
and the internal cylinder must be compressed before gluing because $r_1>r_2$ in
this case.

This problem can be probably solved within the linear elasticity theory though
this solution is unknown to us. Anyway, this solution reproduces only the linear
approximation to the solution obtained within the geometric theory of defects
because the elastic gauge (\ref{efouad}) corresponds to nonlinear elasticity
theory. This example shows that some problems are more easily treated within the
geometric theory of defects.

It is not easy to imagine a space-time as a product of time $t\in\MR$ with the
spacelike cylinder of finite radius $R$ with metric (\ref{econtu}) because it is
geodesically incomplete (a geodesic line reaches the boundary of the cylinder
at a finite value of the proper time). But expression for the metric
(\ref{econtu}) is valid for all values $0\le r<\infty$. Therefore we can
generalize the solution to the whole space-time
\begin{equation*}
  ds^2=dt^2-A^2dr^2-B^2d\vf^2-dz^2,
\end{equation*}
where $t,z\in\MR$, $0\le r<\infty$, and $0\le\vf<2\pi$. It is the solution of
four-dimensional Einstein's equations with energy-momentum tensor (\ref{enmoco})
describing a static conical tube dislocation. This solution is written in the
elastic gauge and therefore depends explicitly on the Poisson ratio $\s$ of
the Universe. If this defect is observed in cosmology, then the Poisson ratio
can be measured. Anyway, the geometric theory of defects provides a way for
measuring the elastic constants of our Universe.
\section{Asymptotically flat wedge dislocation}
\begin{figure}[h,b,t]
\hfill\includegraphics[width=.6\textwidth]{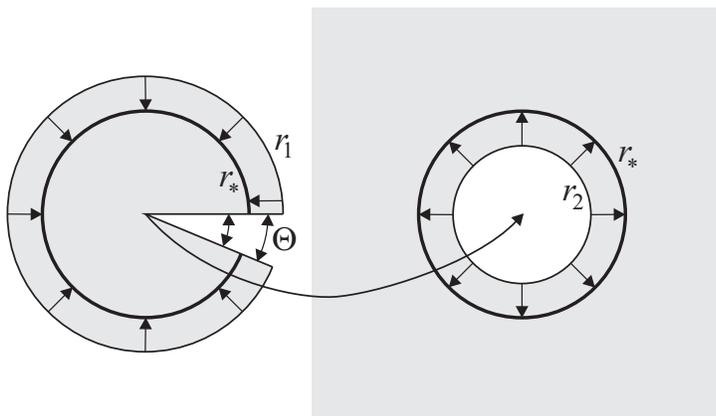}
\hfill {}
\\
\centering \caption{Asymptotically flat conical singularity for negative
deficit angle $\Theta$.}
\label{fasfco}
\end{figure}

The process of defect creation described in the previous section can be
inverted. Suppose that metric has continuous components. Let us take the 
cylindrical rod of media of radius $r_1$ and cut out the wedge of angle $\Theta$ 
from it as shown in Fig.\ref{fasfco}. Then we cut out a cylinder of smaller 
radius $r_2<r_1$ from the infinite media and insert the rod in the media after 
appropriate compression in such a way that circumferences of the internal and 
external cylinders coincide,
\begin{equation}                                                  \label{eqcicu}
  2\pi\al r_1=2\pi r_2,~~~~\al=1+\Theta.
\end{equation}
This guarantees the continuity of the $g_{\vf\vf}$ component of the metric. 
Physically, we have a cosmic string surrounded by the cylindrical shell of
matter in such a way that it becomes flat outside the shell, and we call it
asymptotically flat wedge dislocation.

Let us describe asymptotically flat wedge dislocation mathematically. The source
for the asymptotically flat wedge dislocation differs by the sign from that for
a conical tube dislocation,
\begin{equation}                                                  \label{easflw}
  T_{zz}=-2\Theta\dl(r-r_*).
\end{equation}
Later we shall show that for another choice of the sign in the formula above,
Einstein's equations do not have flat solutions outside the gluing surface at
$r>r_*$. Therefore, solutions with negative deficit angle, $\Theta<0$, have
physical meaning in General Relativity.

Integration of Einstein's equations is similar to the case of the conical tube
dislocation,
\begin{equation*}
  B=-\Theta\int_{r_*}^rdsA(s)+c_1\int_0^rdsA(s)+c_2.
\end{equation*}
We impose the following boundary conditions at the origin and infinity
\begin{equation}                                                  \label{ebouas}
  B_\text{in}|_{r=0}=0,~~~~
  \left.\frac{rA_\text{in}}{B_\text{in}}\right|_{r=0}=\frac{\g_1}\al,
  ~~~~A_\text{ex}|_{r=\infty}=1,
\end{equation}
where $\g_1$ is the positive root of Eq.(\ref{equaga}) and $\al=1+\Theta$. So,
the right hand side of the second equation is written entirely in terms of the
deficit angle $\Theta$ and Poisson ratio. The first two boundary conditions
correspond to a conical singularity at the origin where the metric must have the
form (\ref{easmet}). The third boundary condition corresponds to flat asymptotic
at infinity.

The matching boundary conditions are continuity equations (\ref{emacos}) as for
the conical tube dislocation.

In the internal region,
\begin{align*}
  A_\text{in}&=\frac{B_\text{in}'}{c_1},
\\
  B_\text{in}&=D_1r^{\g_1}+D_2r^{\g_2},~~~~D_{1,2}=\const,
\end{align*}
and the first two boundary conditions in Eq.(\ref{ebouas}) determine three
constants,
\begin{equation*}
  c_1=\al,~~~~c_2=0,~~~~D_2=0.
\end{equation*}

In the external region,
\begin{equation*}
  B'_\text{ex}=-\Theta A_\text{ex}+\al A_\text{ex}=A_\text{ex},
\end{equation*}
and we see the necessity of the minus sign in the source term (\ref{easflw}) for
asymptotic flatness. So, the external solution is
\begin{align*}
  A_\text{ex}&=E_1-\frac{E_2}{r^2},
\\
  B_\text{ex}&=E_1r+\frac{E_2}r,~~~~E_{1,2}=\const.
\end{align*}
The third boundary condition in Eq.(\ref{ebouas}) determines $E_1=1$.

The remaining free integration constants $D_1$ and $E_2$ are found from the
continuity requirements (\ref{emacos}),
\begin{align*}
  D_1r_*^{\g_1}&=r_*+\frac{E_2}{r_*},
\\
  \frac{D_1\g_1r_*^{\g_1-1}}\al&=1-\frac{E_2}{r_*^2}.
\end{align*}
These equations are easily solved,
\begin{align*}
  D_1&=\frac{2\al}{\al+\g_1}r_*^{-\g_1+1},
\\
  E_2&=\frac{\al-\g_1}{\al+\g_1}r_*^2.
\end{align*}

Finally, the essential two dimensional parts of the metric inside and outside 
the gluing surface are
\begin{align}                                                     \label{emetin}
  dl_\text{in}^2&=\frac{4\g_1^2}{(\al+\g_1)^2}\left(\frac r{r_*}\right)^{2\g_1-2}
  \left(dr^2+\frac{\al^2}{\g_1^2}r^2d\vf^2\right),
\\                                                                \label{emetex}
  dl_\text{ex}^2&=
  \left[1-\frac{\al-\g_1}{\al+\g_1}\left(\frac{r_*}r\right)^2\right]^2dr^2+
  \left[1+\frac{\al-\g_1}{\al+\g_1}\left(\frac{r_*}r\right)^2\right]^2r^2d\vf^2.
\end{align}
Their components are continuous functions across the cut, and it is 
asymptotically Euclidean.

Transformation to the Euclidean form of the metric, $dl^2=dy^2+y^2d\vf^2$, is
given by different coordinate changes in the internal and external regions.
Inside the gluing surface,
\begin{equation*}
  y=\frac{2r_*}{\al+\g_1}\left(\frac r{r_*}\right)^{\g_1},~~~~\vf'=\al\vf.
\end{equation*}
So, there is a conical singularity of deficit angle $\Theta$, and
\begin{equation*}
  r_1=\frac{2r_*}{\al+\g_1}.
\end{equation*}
Outside the gluing surface,
\begin{equation*}
  y=r+\frac{\al-\g_1}{\al+\g_1}\frac{r_*^2}r,~~~~\vf'=\vf,
\end{equation*}
and
\begin{equation*}
  r_2=\frac{2\al r_*}{\al+\g_1}.
\end{equation*}
For negative deficit angle, $\Theta<0$, $\al<1$, and $r_2<r_1$ as it should be
from elementary geometric considerations.
\section{Continuous distribution of tube dislocations}
For continuous distribution of tube dislocations, the source term is
\begin{equation}                                                  \label{esotuc}
  T_{zz}=2f(r),
\end{equation}
where $f(r)$ is an arbitrary scalar density of radius. This problem has 
translational symmetry along $z$ axis and circular symmetry in the $r,\vf$ 
plane, as earlier. Einstein's equations (\ref{enmode}) for this source and 
metric (\ref{eunmeg}) reduce to the ordinary differential equation
\begin{equation*}
  \left(\frac{B'}A\right)'=f(r),
\end{equation*}
and can be easily integrated
\begin{equation}                                                  \label{egetub}
  B=\int_0^r dsA(s)\int_0^s dt f(t)+c_1\int_0^r dsA(s)+c_2,~~~~c_{1,2}=\const.
\end{equation}

One constant of integration can be fixed by the requirement that the
circumference of a cylinder surrounding $z$ axis shrinks to zero as $r\to0$.
Then the boundary condition is $B|_{r=0}=0$, and, as the consequence, $c_2=0$.
If the metric is conformally Euclidean at the origin, then the additional
boundary condition is
\begin{equation*}
  \left.\frac{B'}A\right|_{r=0}=1.
\end{equation*}
It fixes $c_1=1$. This consideration clarifies the geometric meaning of the
integration constants. However, one can consider exotic sources (energy-momentum
tensors) with different values of the integration constants $c_{1,2}$.

The elastic gauge for solution (\ref{egetub}) reduces to equation
\begin{equation*}
  \frac{B''}F-\frac{B'f}{F^2}+\frac{B'}{Fr}-\frac B{r^2}+\frac\s{1-2\s}
  \left(\frac{B''}F-\frac{B'f}{F^2}+\frac{B'}r-\frac B{r^2}\right)=0,
\end{equation*}
where
\begin{equation*}
  F(r)=\int_0^r dsf(s)+c_1
\end{equation*}
is a primitive for the source $f(r)$. This equation with appropriate boundary
conditions can be solved at least numerically.

In this way we obtain a solution for an arbitrary continuous distribution of
tube dislocations. This problem shows a great advantage of the geometric theory
of defects as compared to the elasticity theory. Indeed, if the source term
differs from zero everywhere, $f(r)\ne0$, then the curvature tensor is nonzero
due to Einstein's equations, and the displacement vector field does not exist.
This means that we cannot even pose a problem for continuous distribution of
defects within the elasticity theory. Whereas in the geometric theory of
defects, everything is well defined and we can find a metric (i.e.\ elastic
stresses) as a solution of Einstein's equations in the elastic gauge. The
displacement vector field can be introduced only in those domains of space where
the source is zero, $f(r)=0$, (note that in three dimensions, the full curvature
tensor is defined by the Ricci tensor which is zero due to Einstein's equations).
In these domains, the displacement vector field automatically satisfies the
nonlinear elasticity theory equations due to the elastic gauge.
\section{Conclusion }
We have shown that Einstein's equations admit solutions with $\dl$-function type
energy-momentum sources. The components of the metric are not continuous in
general and lead to ambiguous curvature tensor components. Nevertheless, all
ambiguous terms safely cancel in the Einstein's equations. The later reduce to
a linear inhomogeneous equation for a particular combination of metric
components.

The obtained solutions have translational and circular symmetry and correspond
to $\dl$ and $\dl'$ type sources (energy-momentum tensor). In general relativity
and geometric theory of defects, they describe static straight massive thin
cylindrical shells and tube dislocations. In some sense, these are the same
models, because they are governed by Einstein's equations (in the absence of
disclinations). The difference is the elastic gauge which have physical meaning
in the geometric theory of defects and depends explicitly on the Poisson ratio
characterizing elastic properties of media. If we assume that our space-time is
an elastic eather, then the obtained solutions depend explicitly on the Poisson
ratio, and, in principle, we can measure the Poisson ratio of the Universe.

The solution with a $\dl'$ source describes a tube dislocation with
noncontinuous metric component. The corresponding metric is obtained within the
elasticity theory and geometric theory of defects. The results are proved to
coincide. This is a particular case because the elastic gauge corresponds to
nonlinear theory of defects, and, in general, linear elasticity theory
reproduces only the linear approximation of the geometric theory of defects.

The problem with $\dl$-function source has two different types of solutions.
One type is flat inside the gluing surface (massive tube) and conical outside.
The other type of solutions describes conical singularity in the center and
is asymptotically Euclidean. The corresponding solutions within the elasticity
theory are not known. In principle, they can be found but they are hardly
expected to be simple ones because of the complicated boundary conditions. In
these cases Einstein's equations seem to be more easily solved and analyzed.

We also considered the problem of arbitrary continuous distributions of tube
dislocations (massive tubes). This problem is reduced to a linear ordinary
differential equation which can be solved at least numerically. This problem
cannot be even formulated within the elasticity theory because the displacement
vector field does not exist. It demonstrates the advantage of the geometric
theory of defects which is able to describe not only single defects but also
their continuous distribution.

Tube dislocations considered in the present paper may also have applications in
condensed matter physics. For example, they can be used as a continuous model 
for multiwall nanocrystal tubes \cite{deKaKoSh09}.

{\bf Acknowledgments.}
The authors thank I.Shapiro for fruitful comments and discussions. G.B.G.\ would
like to acknowledge financial support from CNPq, FAPEMIG, and FAPES. M.K.\
thanks the Universidade Federal de Juiz de Fora for the hospitality, the
FAPEMIG, the Russian Foundation of Basic Research (Grant No.\ 08-01-00727), and
the Program for Supporting Leading Scientific Schools
(Grant No.\ NSh-3224.2008.1) for financial support.

\end{document}